\documentclass[12pt,preprint]{aastex}
\newcommand{\etal}{\mbox{\rm et al.}}

\newcommand{\lsun}{\mbox{$L_{\odot}$}}
\newcommand{\msun}{\mbox{$M_{\odot}$}}
\newcommand{\rsun}{\mbox{$R_{\odot}$}}
\newcommand{\lstar}{\mbox{$L_{\star}$}}
\newcommand{\mstar}{\mbox{$M_{\star}$}}
\newcommand{\rstar}{\mbox{$R_{\star}$}}

\newcommand{\bc}{\mbox{BC}}
\newcommand{\mbol}{\mbox{$M_{\rm bol}$}}
\newcommand{\mear}{\mbox{$M_\oplus$}}

\newcommand{\vsini}{\mbox{$v \sin i$}}

\newcommand{\teff}{\mbox{$T_{\rm eff}$}}
\newcommand{\fe}{\mbox{\rm [Fe/H]}}
\newcommand{\logg}{\mbox{${\rm \log g}$}}
\newcommand{\giso}{\mbox{$\log g_{\rm iso}$}}
\newcommand{\rhk}{\mbox{$\log R^\prime_{\rm HK}$}}
\newcommand{\shk}{\mbox{$S_{\rm HK}$}}
\newcommand{\prot}{\mbox{$P_{\rm rot}$}}
\newcommand{\mjup}{\mbox{$M_{\rm Jup}$}}
\newcommand{\rchisq}{\mbox{$\chi_{\nu}^2$}}
\newcommand{\Tp}{\mbox{$T_p$}}
\newcommand{\msini}{\mbox{$M\sin i$}}
\newcommand{\ms}{\mbox{m s$^{-1}$}}

\newcommand{\kms}{\mbox{km s$^{-1}$}}
\newcommand{\caii}{\ion{Ca}{2} H \& K}
\newcommand{\alpfe}{\mbox{[$\alpha$/Fe]}}

\received{}
\accepted{}
  
\slugcomment{submitted to ApJ}
  
\shortauthors{Valenti et al.}
\shorttitle{Two Exoplanets Discovered at Keck Observatory}
\begin{document}
  
\title{Two Exoplanets Discovered at Keck Observatory\altaffilmark{1}}
\author{
Jeff A. Valenti\altaffilmark{2},
Debra Fischer\altaffilmark{3},
Geoffrey W. Marcy\altaffilmark{4},
John A. Johnson\altaffilmark{5},
Gregory W. Henry\altaffilmark{6},
Jason T. Wright\altaffilmark{7},
Andrew W. Howard\altaffilmark{4},
Matt Giguere\altaffilmark{3},
Howard Isaacson\altaffilmark{3}
}
  
\email{valenti@stsci.edu}
  
\altaffiltext{1}{Based on observations obtained at the Keck
Observatory, which is operated by the University of California}
  
\altaffiltext{2}{Space Telescope Science Institute,
3700 San Martin Dr., Baltimore, MD 21218}

\altaffiltext{3}{Department of Physics \& Astronomy,
San Francisco State University, San Francisco, CA  94132}
  
\altaffiltext{4}{Department of Astronomy, 
University of California, Berkeley, Berkeley, CA}
  
\altaffiltext{5}{Institute for Astronomy, University of Hawaii,
Honolulu, HI 96822}

\altaffiltext{6}{Center of Excellence in Information Systems,
Tennessee State University, 3500 John A.\ Merritt Boulevard,
Box 9501, Nashville, TN 37209} 
  
\altaffiltext{7}{Department of Astronomy, 226 Space Sciences
Building, Cornell University, Ithaca, NY 14853}

\begin{abstract}
We present two exoplanets detected at Keck Observatory. HD 179079
is a G5 subgiant that hosts a hot Neptune planet 
with $\msini=27.5$ \mear\ in a 14.48 d,
low-eccentricity orbit. The stellar reflex velocity induced 
by this planet has a semiamplitude of $K = 6.6$ \ms. 
HD 73534 is a G5 subgiant with a Jupiter-like
planet of $\msini=1.1$ \mjup\ and $K = 16$ \ms\ in a nearly circular 
4.85 yr orbit. Both stars are chromospherically inactive and metal-rich. 
We discuss a known, classical bias in measuring eccentricities for 
orbits with velocity semiamplitudes, $K$, comparable to the radial 
velocity uncertainties. For exoplanets with periods longer than 10 days, the 
observed exoplanet eccentricity distribution is nearly flat for 
large amplitude systems ($K>80$ \ms), but rises linearly toward 
low eccentricity for lower amplitude systems ($K>20$ \ms).
\end{abstract}


\section{Introduction}
Since the discovery of 51 Peg \citep{may95}, more than 300
exoplanets have been detected, mostly by Doppler measurements
of stellar reflex velocities. The distributions of these exoplanet 
masses, semi-major axes, and orbital eccentricities provide 
evidence for planet formation and orbital evolution \citep{mar08}.
Currently, known exoplanets
have a median mass of about 1 \mjup\ and median semi-major
axis of about 1 AU. Early exoplanet discoveries were mostly
massive gas giants in short-period orbits because such orbits
have velocity amplitudes much larger than measurement errors
and because it takes less time to observe many orbits and get
complete phase coverage. Steady improvements in Doppler
precision have enabled the recent detection of planets
with $\msini\sim10\mear$ \citep{how09,riv05,udr07,may09},
despite velocity semiamplitudes of only a few \ms.

Here, we present two new exoplanets detected at Keck Observatory as
part of a search for hot Neptune-mass and other low-amplitude planets.
The host stars were originally observed as part of the N2K program,
a survey of metal-rich stars to detect hot Jupiters \citep{fis05a}.
We continued observing a subset of promising N2K stars to search for
exoplanets with lower velocity amplitudes, including Neptune mass
planets in short period orbits (``hot Neptunes''). The hot Neptune
sample consists of about a hundred N2K stars with low chromospheric
activity, low \vsini, and a velocity scatter greater than
$2\sigma$ but less than 20 \ms. The hot Neptune sample inherits
from the N2K survey a selection bias in favor of high-metallicity
stars that increases the probability of detecting massive planets
\citep{fis05b} and a Malmquist bias that increases the number of
subgiants because stellar magnitude was a factor in target selection.

The orbital eccentricity distribution is an interesting characteristic
of detected exoplanets. In sharp contrast to planets in our solar
system, exoplanets with orbital periods longer than 10 d have
eccentricities that range from circular to greater than 0.9,
with a median eccentricity of 0.24. Planets with orbital periods
shorter than about 10 d are expected to circularize over time via
tidal interactions with the host star. Consistent with this prediction,
the median eccentricity of exoplanets with orbital periods shorter
than 10 days is only 0.013. Interesting exceptions include HD 185269b
\citep[][$P=6.84$ d, $e=0.30$]{joh06}, HD 147506b \citep[][$P=5.6$ d,
$e=0.5$]{bak07}, and \citep[][$P=3.2$ d, $e=0.26$]{joh08}. 
Precise eccentricity measurements for planets with a range of periods, masses,
and ages help to empirically constrain orbital evolution models.

\section{Data and Methods\label{sec_methods}}
\subsection{Spectroscopic Observations}

We used the HIRES spectrometer \citep{vog94} at Keck Observatory for
3--5 years to obtain a temporal sequence of $R=65,000$ spectra for each
star. An iodine cell in the beam imprinted a rich set of molecular
absorption lines on the stellar spectrum. These iodine lines constrain
the wavelength scale, point-spread functions, and Doppler shift for
each individual observation \citep{mar92,but96}. An exposure meter
was used to adjust each exposure time to achieve a consistent
signal-to-noise ratio of 200 per extracted pixel, which alleviates
some types of systematic errors. The exposure meter was also used
to determine the photon-weighted midpoint of each exposure, which
improves the precision of the barycentric velocity correction.

Spectra obtained after 2004 August with the new HIRES detector mosaic
include the \caii\ lines, which provide a diagnostic of chromospheric
activity. We characterized line core emission in terms of the S index
\citep{vau78,dun91}. To improve measurement precision, we matched the
line wings and neighboring continua for each observation of a given
star to the mean value for all observations of that star \citep{isa09}.

\subsection{Photometric Observations\label{sec_photometry}}

We obtained photometric observations of HD~179079 and HD~73534
with the T12 0.8 m automated photometric telescope (APT) at Fairborn
Observatory in southern Arizona.  The T12 APT and its precision
photometer are very similar to the T8 APT described in \citet{hen99}.
The precision photometer uses two temperature-stabilized EMI 9124QB
photomultiplier tubes to measure photon count rates simultaneously
through Str\"omgren $b$ and $y$ filters. 

The telescope was programmed to measure each target star with
respect to three nearby comparison stars in the following
sequence: DARK, A, B, C, D, A, SKY$_{\rm A}$, B, SKY$_{\rm B}$,
C, SKY$_{\rm C}$, D, SKY$_{\rm D}$, A, B, C, D, where A, B, and
C are comparison stars and D is the program star. Each complete
sequence, referred to as a group observation, was reduced to
form three independent measures of each of the six differential
magnitudes D$-$A, D$-$B, D$-$C, C$-$A, C$-$B, and B$-$A.
The differential magnitudes were corrected for differential
extinction with nightly extinction coefficients and transformed
to the standard Str\"omgren system with yearly mean
transformation coefficients. To filter out observations taken
under non-photometric conditions, an entire group of observation
was discarded if the standard deviation of any of the six mean
differential magnitudes exceeded 0.01 mag. We also combined
the Str\"omgren $b$ and $y$ differential magnitudes into a
single $(b+y)/2$ passband to improve the precision.

\subsection{Stellar Analysis\label{sec_stellar}}

We determined stellar properties by analyzing one spectrum
of each star obtained without the iodine cell. We used SME
\citep{val96} and the procedure of \citet{val05} to fit each
observed spectrum with a synthetic spectrum, obtaining the
stellar effective temperature \teff, surface gravity \logg,
metallicity [M/H], projected rotational velocity \vsini,
and elemental abundances of Na, Si, Ti, Fe, and Ni. Our [M/H]
parameter here \citep[as in][]{val05} scales solar abundances
for elements other than Na, Si, Ti, Fe, and Ni that have
significant spectral lines in our fitted wavelength intervals.
We report iron abundance \fe\ rather than [M/H] because \fe\
is better defined. We use the abundance of Si relative to
iron [Si/Fe] as a proxy for alpha-element enrichment \alpfe.

We obtained a bolometric correction (\bc) for each star by
interpolating the ``high-temperature'' grid of \citet{van03}
to our spectroscopically determined \teff, \logg, and \fe.
Using formulae and constants in \citet{val05}, we calculated
stellar luminosity from apparent visual magnitude $V$, BC,
and distance $d$ of the star, and then stellar radius \rstar\
from \lstar\ and \teff.

Following \citet{val05}, we determined stellar mass \mstar,
age, and surface gravity \giso\ by interpolating tabulated
Yonsei--Yale isochrones \citep{dem04}, first to our measured
\alpfe, then to our measured \fe, then to our measured \teff,
and finally to the observed \lstar. This yields zero, one,
or multiple possible stellar models. We repeat the
interpolation for a grid of $17^4$ values of \alpfe, \fe,
\teff, and \lstar\ that span the range of measurement
uncertainties. We weight each outcome by the likelihood
of the interpolation parameter and by the lifetime of the
output evolutionary state. This favors relatively stable
main-sequence states over rapidly evolving states.

\begin{figure}
\plotone{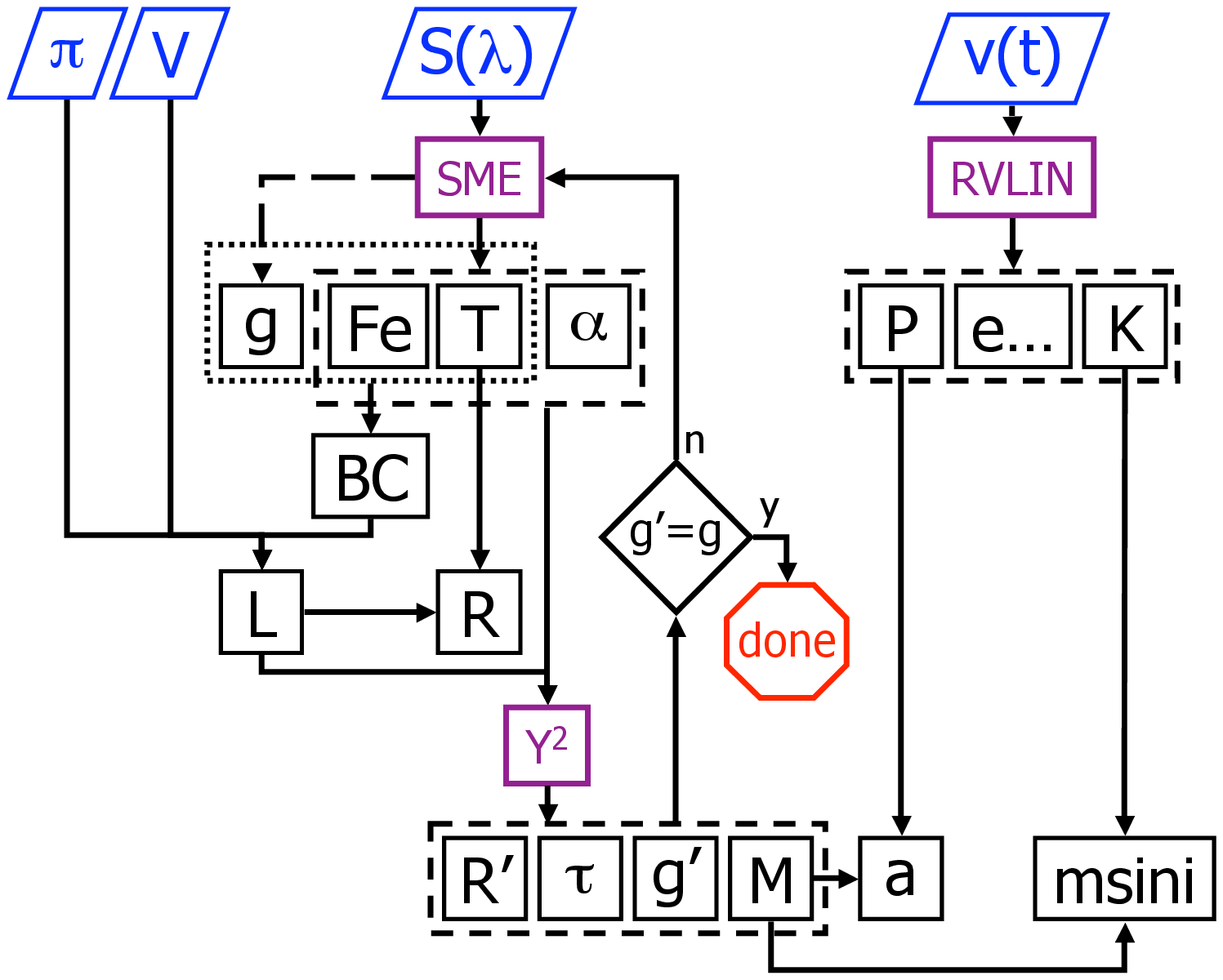}
\figcaption{Graphical representation of the analysis flow.
Observables in the top row are parallax, apparent visual
magnitude, a high-resolution spectrum, and measured radial
velocities. The left side shows the spectroscopic (SME)
and isochrone (Y$^2$) analysis, while the right side shows
the orbital analysis (RVLIN). Symbols for derived quantities
are as described in the text, but with Fe for \fe,
$\alpha$ for \alpfe, $\tau$ for age, and $g^\prime$ for
$g_{\rm iso}$. The arrow pointing up to the $g^\prime=
g_{\rm iso}$ decision diamond illustrates a new outer loop
that enforces consistency between spectroscopic and
isochrone gravities. (A color version of this figure is available 
in the online journal.)
\label{fig_analysis_flow}}
\end{figure}

The \citet{val05} procedure described above does not require
agreement between \logg\ from spectroscopy and \giso\ from
isochrones, though the two are usually close.
Here we introduce an outer iteration loop in which the SME
analysis is repeated with \logg\ fixed at the value of \giso\
from the preceding iteration. After a few iterations,
the output \logg\ from SME agrees with the input \giso.
The price for this self-consistency is a greater dependence
on models and worse \rchisq\ for the spectrum fits. However,
we are already completely dependent on stellar evolution
models for \mstar, and systematic errors in spectral line data
dominate \rchisq. Using evolutionary tracks as an additional
constraint on \logg\ while fitting spectra should improve the
accuracy of the other derived parameters. We give examples
later in this paper. Figure \ref{fig_analysis_flow} shows
the analysis procedure graphically.

\subsection{Orbital Analysis\label{sec_orbital}}

Using the periodogram procedure described in \citet{mar05}, we
analyzed the velocity time series for each star to identify planet
candidates and prospective periods. Once a sufficient number of
observations were obtained, we fitted the velocities with Keplerian
orbits, using a Levenberg-Marquardt algorithm \citep{wri09}.
The free parameters are orbital period $P$, velocity semiamplitude
$K$, eccentricity $e$, time of periastron passage $T_p$, argument
of periastron referenced to the line of nodes $\omega$, and velocity
offset of the center-of-mass $\gamma$.

When fitting Keplerian models, we adopted a total variance
equal to the quadrature sum of our velocity measurement
precision (1--2 \ms) and a velocity jitter term of 3.5 \ms.
Jitter includes both systematic measurement errors and
astrophysical velocity perturbations caused by photospheric 
flows and inhomogeneities. Our adopted jitter of 3.5 \ms\
for the two subgiants in this paper yields \rchisq\ near unity
and is consistent with the range of values in \citet{wri05}.
The exact choice of jitter has an insignificant effect on our
derived values of $P$ and $K$.

We used $10^5$ Monte Carlo trials to estimate uncertainties in our
derived orbital parameters. In each trial, we constructed a simulated
observation by adding errors to the Keplerian model that best matches
our observed velocities. These errors were selected randomly from
the distribution of observed residuals. A particular residual could be
used multiple times or not at all in any given trial. We fitted each
simulated observation using exactly the same procedure that we used
to analyze the actual observation. Each trial included a periodogram
analysis, which for stars with few observations can occasionally
yield a distinct period. After $10^5$ trials, the width of the
distribution function for a particular orbital parameter yields
an estimate of the uncertainty in our observed value of that
parameter.

\section{HD 179079}
\subsection{Stellar Characteristics}

HD 179079 (HIP 94256, $V=7.95$, $B-V=0.744\pm0.013$) is a G5
subgiant at a distance of $65.5\pm3.3$ pc \citep{per97,van08}.
High-resolution spectroscopic analysis (see Section 2.3)
yields $\teff=5684\pm44$ K, $\logg=4.06\pm0.06$, $\fe=+0.25\pm0.03$
dex, and $\vsini<1.0$ \kms. These spectroscopic results imply a
bolometric correction of $\bc=-0.094$ and hence a stellar
luminosity of $\lstar=2.41\pm0.27$ \lsun, where the uncertainty
in luminosity is dominated by the uncertainty in distance.
Using \lstar\ and \teff, we obtain a stellar radius of
$\rstar=1.60\pm0.09$ \rsun. Stellar evolutionary
tracks imply a stellar mass of $\mstar=1.15\pm0.03$ \msun.

HD 179079 is chromospherically inactive, based on weak emission
in the cores of the \caii\ lines. We measure \shk = 0.153, which
yields $\rhk=-5.06$ and implies a rotation period of roughly 38 d
\citep{noy84}. The level of chromospheric activity implies an
approximate age of 7 Gyr, which is consistent with the 68\%
credible interval of 6.1--7.5 Gyr from the isochrone analysis.
We estimate that surface motion in the photosphere of this early
subgiant contributes 3.5 \ms\ of velocity jitter that we add in
quadrature to velocity measurement uncertainties, when modeling
the data. Table \ref{tab_stellar_params} summarizes the stellar
parameters. Table \ref{tab_179079_iter} shows how key stellar
parameters changed during the iteration procedure introduced
in Section 2.3.

\subsection{Doppler Observations and Keplerian Fit}

We obtained 74 observations of HD 179079 beginning in 2004 July. 
The observation dates, radial velocities and measurement
uncertainties are listed in Table \ref{tab_rv_179079}.
Typical exposure times were about 2 minutes. The median
velocity measurement uncertainty is 1.2 \ms, which is small
compared to the velocity jitter of 3.5 \ms.

\begin{figure}
\plotone{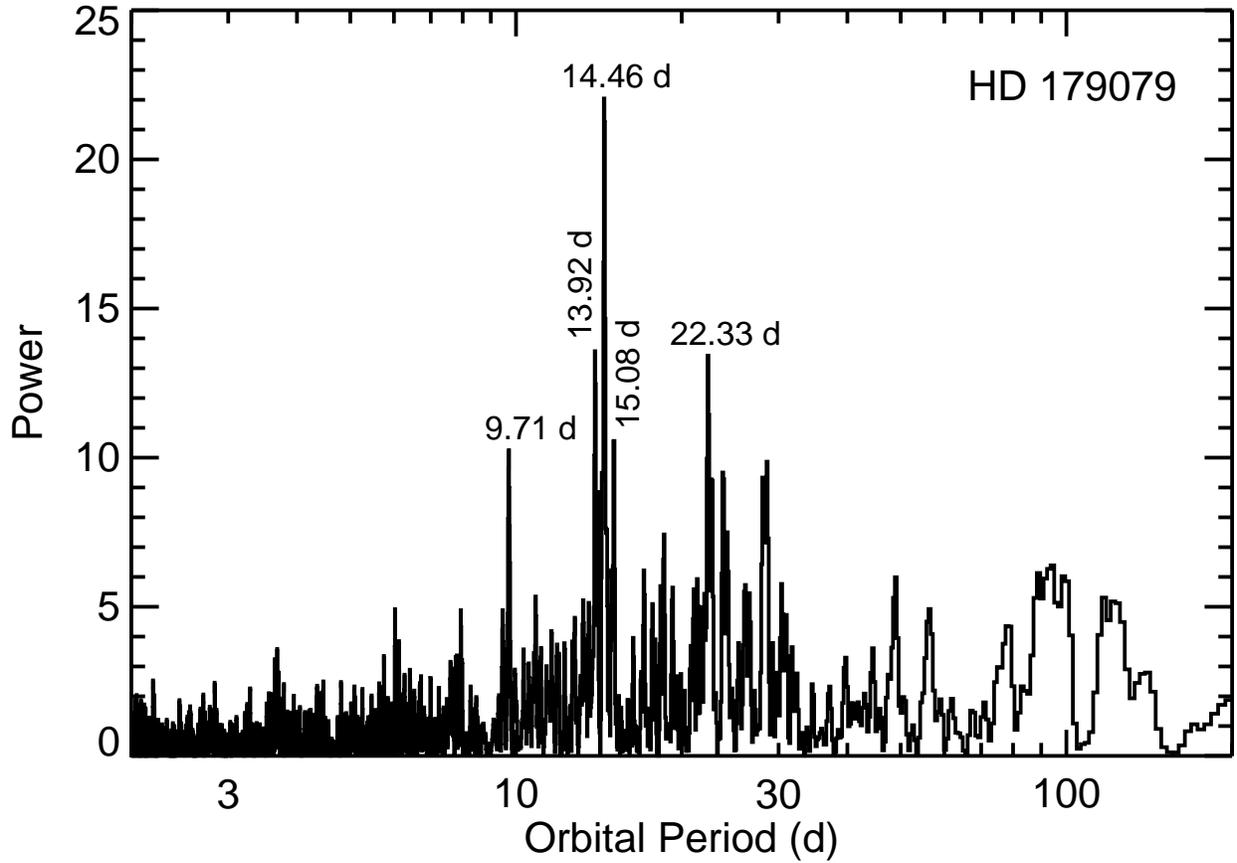}
\figcaption{Periodogram of 74 radial velocity measurements
for HD 179079. The prominent peak at 14.46 d is consistent
with the $14.476\pm11$ d period returned by a Keplerian fit
of the measured velocities. The next four strongest peaks
(labeled in the figure) are significantly weaker.
\label{fig_power179079}}
\end{figure}

Figure \ref{fig_power179079} shows a periodogram of the 74
measured radial velocities, with an unambiguous peak in power
at 14.46 days. The false alarm probability (FAP) is less than
0.0001, i.e., the probability that a random set of data would
produce a peak with the observed power is less than 0.01\%.
The FAP test checks for spurious peaks
that can arise because of window functions in the data. To
calculate the FAP, simulated velocities for actual observation
times were drawn randomly from the distribution of observed
velocities, allowing reuse of any value. After $10^4$ trials,
we found no peaks for simulated data with as much power as the
observed peak at 14.48 days. Thus, the FAP is less than the
reciprocal of the number of trials, i.e. less than $10^{-4}$.

Our best-fitting Keplerian model has a period $P=14.476\pm0.011$ d,
velocity semiamplitude $K=6.64\pm0.60$ \ms, and eccentricity 
$e=0.12\pm0.09$. Uncertainties for the orbital parameters were
derived from Monte Carlo trials, as described in Section 2.4.
The rms residual about the model fit is 3.9 \ms. Including jitter
of 3.5 \ms, we find $\rchisq=1.04$. The eccentricity is poorly
constrained, as indicated by the large uncertainty found in our
Monte Carlo trials. A circular orbit yields $\rchisq=1.03$,
which is as probable as the solution obtained with eccentricity
and $T_p$ as free parameters. The challenge of measuring low
eccentricities in low amplitude systems is discussed in Section 5.1. 

\begin{figure}
\plotone{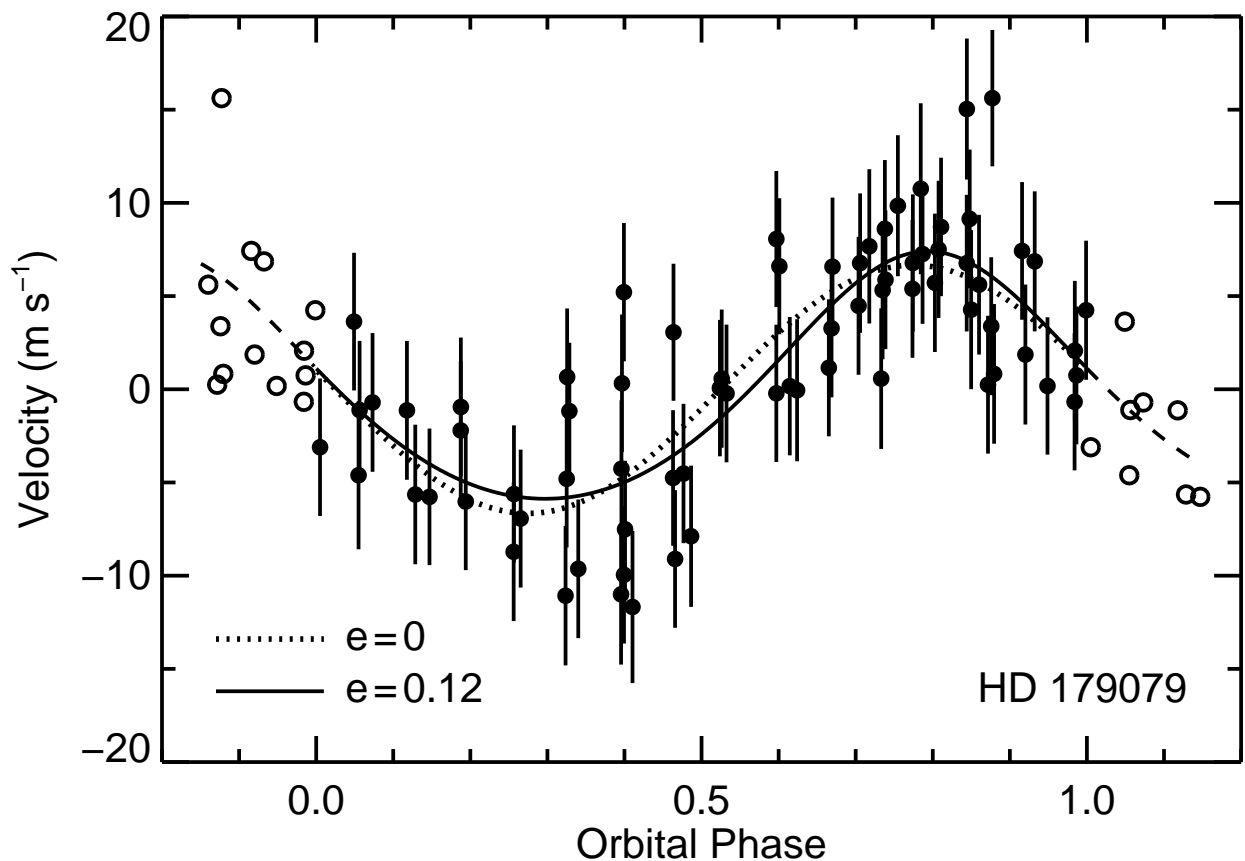}
\figcaption{Phased radial velocities for HD 179079 reveal an orbital
period of 14.48 d, a velocity amplitude of 6.64 \ms, and an eccentricity 
of 0.12. Error bars illustrate the quadrature sum of the velocity precision
for each measurement and 3.5 \ms\ of jitter (systematic errors and/or
intrinsic stellar variability). The Keplerian model is overplotted with a
solid line  and the dotted line shows the model with eccentricity fixed
to zero.  Adopting a stellar mass of 1.15 \msun\ we derive a planet
mass,  $\msini=28$ \mear\ and orbital radius of 0.122 AU. 
\label{fig_rvfit_179079}}
\end{figure}

Figure \ref{fig_rvfit_179079} shows the phase-folded RV data
together with the Keplerian (solid line) and circular (dotted
line) models that best fit the data. Observations are plotted
using phases based on the Keplerian fit, rather than the
circular fit. In this and subsequent velocity plots, error
bars are dominated by velocity jitter, rather than velocity
measurement precision. Adopting a stellar mass of $1.146\pm0.028$
\msun, we derive a minimum mass of $\msini=27.5\pm2.5$ \mear\
and a semi-major axis of $0.121\pm0.001$ AU. The complete
Keplerian orbital solution is given in Table
\ref{tab_orbital_params}.
  
\subsection{Transit Search}
Because the orbital period for this planet is relatively short, we
have carried out an extensive search for transits, both with
photometry (described  in the following section) and by phase-folding
the radial velocities to search for velocities that might have
been obtained serendipitously during the $\sim3$ hr transit window.
As the planet traverses the stellar disk, it blocks light from first
one side (say the approaching, blueshifted edge of the stellar disk)
and then the other side (the receding, redshifted edge) of the star.
Our Doppler analysis interprets the spectral line asymmetry as excess
velocity shifts during ingress and egress. This phenomenon is known as
the Rossiter-McLaughlin effect.

The amplitude of the Rossiter-McLaughlin effect increases with both
projected stellar rotation velocity and planet size relative to the
star. HD 179079 has a low rotational velocity ($\vsini=0.5$ \kms)
and a small radius (expected to be similar to Neptune for an
$\msini = 25.4$ \mjup\ planet). Therefore, the amplitude of the
Rossiter-McLaughlin effect is expected to be no more than our
measurement errors of about 2 \ms\ for HD 179079.
Nevertheless, we phase-folded the radial velocities and calculated
the ingress and egress times for eccentricities ranging from 0 to 0.1.
Uncertainties in the orbital eccentricity lead to shifts of up to
16 hr in the prospective transit time at the epochs of our radial
velocity measurements. Velocity residuals during these broad transits
windows were no larger than velocity residuals at other orbital phases.

\subsection{Photometry}

Our 243 good brightness measurements of HD 179079 were made between
2007 June  and 2008 June and cover parts of the 2007 and 2008
observing seasons. The comparison stars A, B, and C were HD 177552
($V=6.51$, $B-V=0.36$, F1~V), HD 181420 ($V=6.57$, $B-V=0.44$, F2),
and HD 180086 ($V=6.63$, $B-V=0.35$, F0), respectively.
The differential magnitudes C$-$A, C$-$B, and B$-$A demonstrated
that all three comparison stars were constant to 0.002 mag or better.
To minimize the effect of any low-level intrinsic variation in the
three comparison stars, we averaged the three D$-$A, D$-$B, and
D$-$C differential magnitudes of HD 179079 within each group into
a single value, representing the difference in brightness between
HD 179079 and the mean brightness of the three comparison stars:
$D-(A+B+C)/3$. The standard deviation of these ensemble differential
magnitudes for the complete data set is 0.00190 mag. This is
comparable to the typical precision of a single observation with
this telescope, indicating there is little or no photometric 
variability in HD 179079.

Solar-type stars often exhibit brightness variations caused by cool,
dark photospheric spots as they are carried  into and out of view
by stellar rotation \citep[e.g.,][]{gai00}. Periodogram analyses of
the D$-$A, D$-$B, and D$-$C differential magnitudes for HD 179079
yield no significant periodicity between 1 and 100 days, consistent
with the star's low level of chromospheric activity and its low
\vsini\ (Table \ref{tab_stellar_params}). We see no significant
power at any period within a factor of 2 of 38 d, which is the
rough rotation period implied by the chromospheric activity level.

\begin{figure}
\plotone{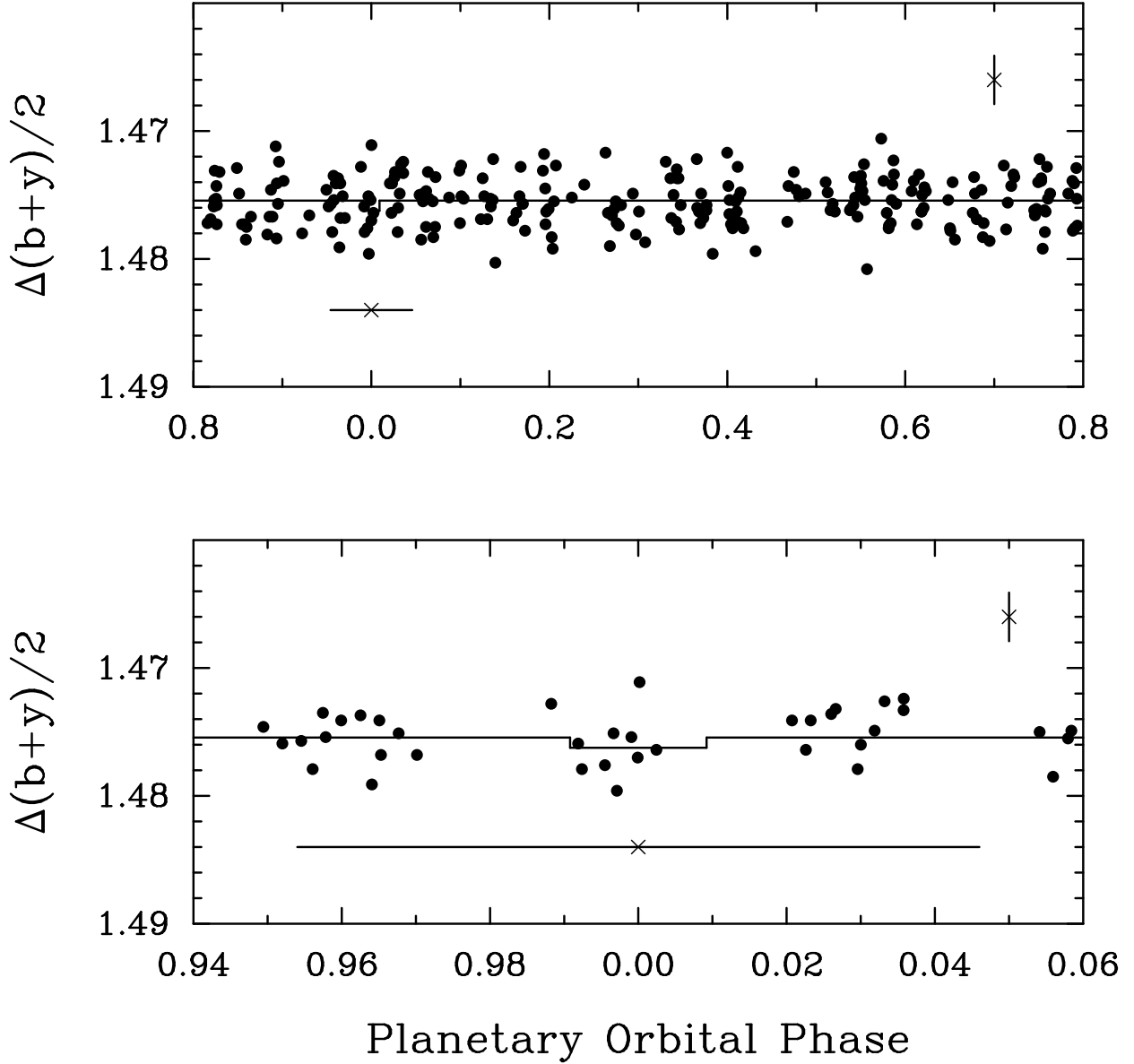}
\figcaption{
\textit{Top panel:} the 243 ensemble $D-(A+B+C)/3$ photometric 
observations of HD 179079 in the combined Str\"omgren $(b+y)/2$
passband, acquired with the T12 0.8 m APT over two observing
seasons and plotted modulo the 14.476 d orbital period of the
inner companion. Phase 0.0 corresponds to a predicted time of
mid transit. A least-squares sine fit at the orbital period
yields a semiamplitude of only $0.00007\pm0.00016$ mag.
\textit{Bottom panel:} the photometric observations of HD 179079
near the predicted time of transit plotted with an expanded
scale on the abscissa. The solid curve shows the predicted
time of transit with a drop in stellar brightness of 0.08\%
($\sim0.00087$ mag). The error bar in the upper right  of both
panels represents the mean precision of a single observation
(0.0019 mag). The error bar immediately below the predicted
time of transit in both panels represents the uncertainty
in the predicted time of mid-transit ($\pm0.67$ days or
$\pm0.046$ phase units).
\label{fig_phot_179079}}
\end{figure}

The 243 ensemble $(b+y)/2$ differential magnitudes of HD 179079
are plotted in the top panel of Figure \ref{fig_phot_179079}.
Phases are computed from the orbital period given in Table
\ref{tab_orbital_params} and the epoch JD $2,454,678.8\pm0.67$,
a recent time of mid-transit derived from the orbital elements.
A least-squares sine fit on the orbital period yields a
semiamplitude of only 0.00007 $\pm$ 0.00016 mag.  
This very low limit to photometric variability on the radial
velocity period is strong evidence that the low-amplitude
radial velocity variations observed in the star are, in fact,
due to reflex motion induced by a low-mass companion and not
to activity-induced intrinsic variations in the star itself 
\citep[e.g.,][]{pau04}.  

The photometric observations of HD 179079 near the predicted time
of transit are replotted with an expanded horizontal scale in the
bottom panel of Figure \ref{fig_phot_179079}. The solid curve shows
the predicted time (0.00 phase units) and duration ($\pm0.009$
phase units) of transits with a depth of 0.08\% computed from
estimated stellar and planetary radii. The error bar in the upper
right of both panels represents the mean precision of a single 
observation (0.0019 mag).  The horizontal error bar immediately
below  the transit in both panels represents the uncertainty in
the predicted time of mid-transit ($\pm0.67$ days or $\pm0.046$
phase units). It is clear from the data in the bottom panel that
we cannot rule out the possibility of shallow transits of
HD 179079b.

\section{HD 73534}
\subsection{Stellar Characteristics}

HD 73534 (HIP 42446, $V=8.23$, $B-V=0.962\pm0.021$) is a G5
subgiant at a distance of $81.0\pm4.9$ pc \citep{esa97,van08}.
High-resolution spectrum synthesis modeling yields
$\teff=5041\pm44$ K, $\logg=3.78\pm0.06$, $\fe=+0.23\pm0.03$
dex, and $\vsini<1.0$ \kms. The resulting bolometric correction
of $\bc=-0.266$ yields a stellar luminosity of
$\lstar=3.33\pm0.43$ \lsun. Combining \teff\ and \lstar, we
obtain a stellar radius of $\rstar=2.39\pm0.16$ \rsun. Stellar
evolutionary tracks yield a stellar mass of $\msun=1.23\pm0.06$
\msun.

HD 73534 has minimal emission in the \caii\ line cores,
implying chromospheric inactivity, which is typical for
subgiants \citep{wri04}. Applying \citet{noy84} relationships
that were calibrated using main-sequence stars, our measured
$\shk = 0.155$ yields $\rhk=-5.13$ and a crude rotation
period of 53 d. The level of chromospheric activity
implies an approximate age of 9 Gyr, which differs
significantly from the 68\% credible interval of 5.2--7.2 Gyr
from the isochrone analysis.

Subgiants have slightly more velocity jitter than inactive
main-sequence stars, as evidenced by the greater rms scatter
seen in subgiants without detected planets. We estimate that
the intrinsic stellar jitter of HD 73534 is 3.5 \ms.
Table \ref{tab_stellar_params} lists our derived stellar
parameters. Table \ref{tab_73534_iter} shows how key
parameters for HD 73435 changed during the iteration
procedure introduced in Section 2.3.
 
\subsection{Doppler Observations and Keplerian Fit}
  
We began observing HD 73534 in 2004 July as part of the N2K
program \citep{fis05a}. No short-period velocity variations
were detected, but we continued to obtain a few velocity
measurements each year to map out an emerging low-amplitude,
long-period planet. We now have a total of 30 observations
that span five years. The raw velocity measurements have a
precision of 1.1 \ms, but the (unmodeled) rms is 13.8 \ms.
Exposure times were two to five minutes. The observation
dates, radial velocities and associated uncertainties are
listed in Table \ref{tab_rv_73534}. 

\begin{figure}
\plotone{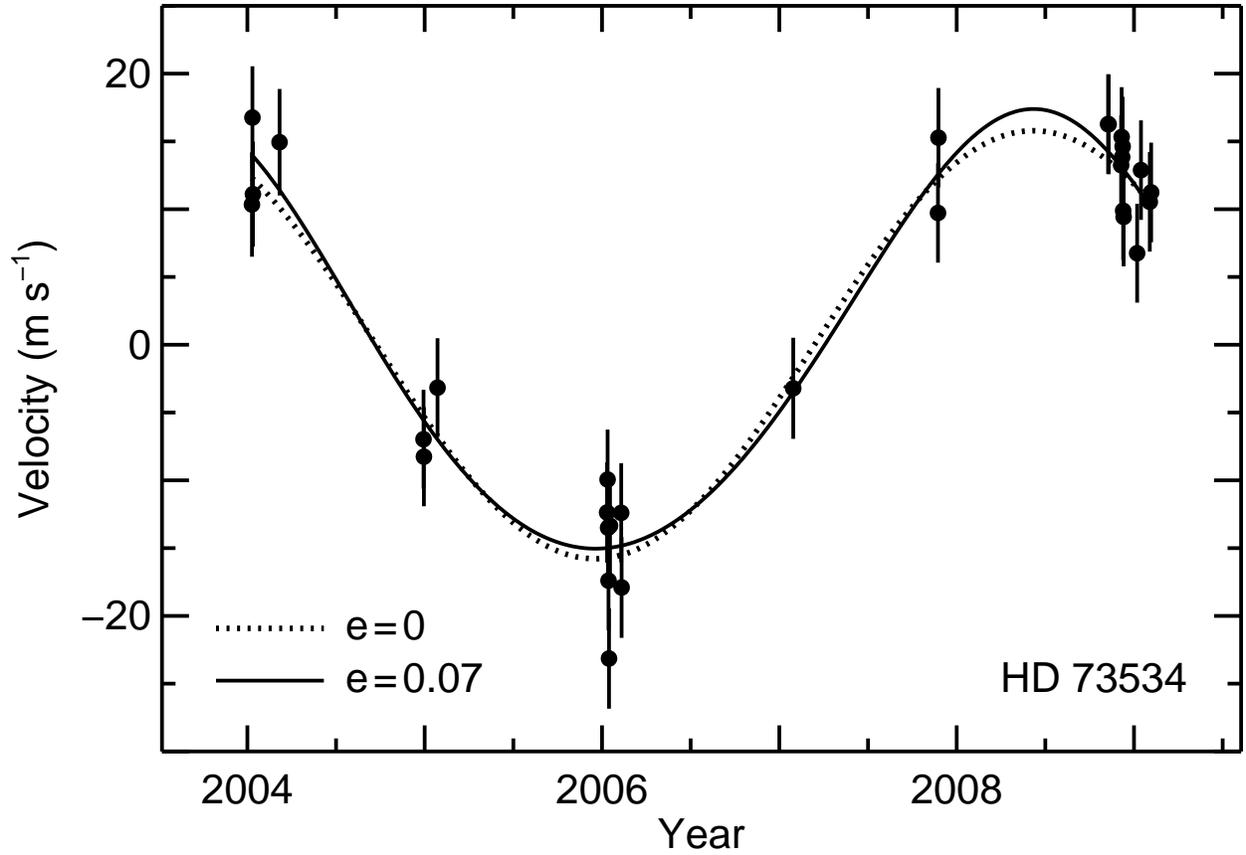}
\figcaption{Radial velocities for HD 73534
are fit with an orbital period of 1770 d, velocity amplitude 
of 16.2 \ms, and a nearly circular orbit. Error bars illustrate
the quadrature sum of the velocity precision for each
measurement and 3.5 \ms\ of jitter (systematic errors
and/or intrinsic stellar variability). The assumed stellar
mass of 1.23 \msun\ yields $\msini=1.10$ \mjup\ and
a semi-major axis of 3.07 AU.
\label{fig_rvfit_73534}}
\end{figure}

The best fit Keplerian model has a period, $P=1770\pm40$ d,
velocity semiamplitude, and $K=16.2\pm1.1$ \ms. The orbital
eccentricity, $e=0.07\pm0.07$, is not significantly different
from zero. The rms residual of the fit is 3.36 \ms\ with
\rchisq = 0.91 after including stellar jitter of 3.5 \ms.
Adopting a stellar mass of $\msun=1.228\pm0.060$ \msun, we
derive $\msini=1.103\pm0.087$ \mjup\ and a semimajor axis of
$a=3.067\pm0.068$ AU. The complete set of orbital parameters
are listed in Table \ref{tab_orbital_params}. Figure
\ref{fig_rvfit_73534} shows the phased radial velocity data
with the best-fit Keplerian (solid line) and circular
(dotted line) models overplotted.

\subsection{Photometry}

From 2004 November to 2008 December, we collected 521 good
photometric observations of HD 73534 during five consecutive
observing seasons. We do not see a correlation between radial 
velocities and activity or photometric measurements, 
however, we present the photometric data for posterity. 

The comparison stars A, B, and C were
HD 72943 ($V=6.33$, $B-V=0.34$, F0 IV), HD 73347 ($V=8.00$,
$B-V=0.40$, F0), and HD~73821 ($V=7.82$, $B-V=0.30$, F0),
respectively. Comparison star A (HD 72943) was found to be
variable with an amplitude of 0.010 mag and a period of
0.0919 d; it is probably a $\delta$~Scuti variable.
Comparison stars B and C were constant to 0.002 mag or
better, so we created and analyzed ensemble differential
magnitudes using only those two comparison stars: $D-(B+C)/2$.

The standard deviation of all 521 observations is 0.0018 mag,
which is the typical precision of our photometry. To search
for low-amplitude periodic brightness variations, we calculated
power spectra for each observing season and combined the results.
We did not detect any significant power for periods between
1 and 100 days, which more than spans the range of plausible
rotation periods. The absence of significant optical variability
on rotational time scales is consistent with the low level of
chromospheric activity detected in HD~73534.

To search for low-amplitude, brightness variations on longer
time scales, we computed the mean brightness for each of our
five observing seasons. For HD~73534, the seasonal means have
a full range of 0.0029 mag and a standard deviation of 0.0012
mag. For the two comparison stars, the seasonal mean values
of $C-B$ have a full range of only 0.0014 mag and a standard
deviation of only 0.0006 mag, a factor of 2 smaller.
\citet{wri08} demonstrated that for solar-type stars, we can
measure seasonal means with a precision (standard deviation)
of 0.0002 mag. HD~73534 varies by about 0.001 mag from year
to year, but the variations are not systematic.

Finally, we compared our seasonal mean brightness measurements
with our seasonal mean radial velocities, obtaining a linear
correlation coefficient of 0.3345. With only five data points,
the correlation coefficient would be at least this large 42\%
of the time, if brightness and velocity are uncorrelated
random variables. A significant correlation between stellar
brightness and radial velocity would raise doubts about the
planetary origin of the stellar velocity variations, but in
this case we do not detect a significant correlation.

\section{Eccentricity\label{sec_ecc}}
\subsection{Eccentricity Measurement Bias\label{sec_ebias}}

Eccentricity cannot be negative. For a circular orbit, errors
in individual radial velocity measurements can only drive the
measured eccentricity away from the true value of zero
\citep[e.g.,][]{she08}.
As radial velocity planet searches push to lower amplitude
systems, this bias becomes significant for low eccentricity
planets.

When observed velocity constraints for a circular orbit are
uniformly distributed in orbital phase (approximately true
for most radial velocity planet detections), \citet{luc71}
derives the probability distribution for measured eccentricity,
\begin{equation}
p(e)de={e\over\sigma_e^2}\exp\left(-{e^2\over2\sigma_e^2}\right)de,
\label{eq_edist}
\end{equation}
where the eccentricity uncertainty, $\sigma_e$, is given by
\citet{luy36} as
\begin{equation}
\sigma_e={\sigma\over K}\left({2\over N}\right)^{0.5},
\label{eq_sigmae}
\end{equation}
where $\sigma$ is the typical uncertainty in velocity and $N$
is the number of velocity measurements.
The detection of nonzero eccentricity with better than 95\%
confidence requires approximately $e/\sigma_e>2.45$.

\begin{figure}
\plotone{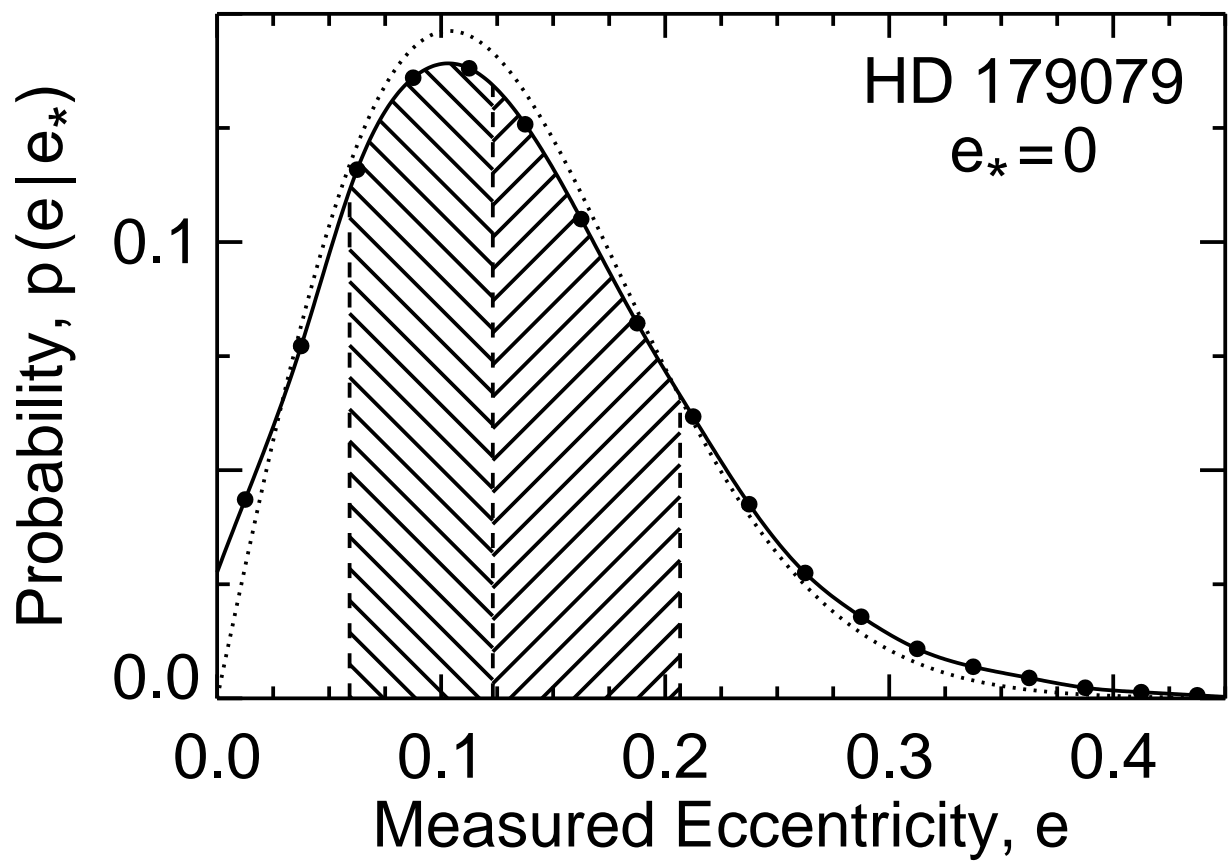}
\figcaption{Probability that we would measure an eccentricity
$e$ for our 74 Keck observations of HD 179079, if the true
eccentricity is zero, $e_*=0$.
The asymmetric distribution of measured eccentricities has a
median of 0.123 and a standard deviation of 0.075, despite a
true eccentricity of zero. The hatched regions show 34.1\% of
the distribution on either side of the median.
\label{fig_ebias}}
\end{figure}

To demonstrate the challenge in recovering zero eccentricity 
for HD 179079, we created $10^5$ synthetic data sets based on
an $e=0$ fit of the observed velocities for HD 179079. We added
errors by drawing randomly (with replacement) from the observed
distribution of residuals about the $e=0$ fit. Finally, we fitted
each simulated data set with a Keplerian, leaving eccentricity
as a free parameter. Figure \ref{fig_ebias} shows the resulting
distribution of imprecisely measured eccentricities.

Even though we simulated a circular orbit, the measured
eccentricities in Figure \ref{fig_ebias} are significantly
biased toward positive values. The eccentricity distribution
has a median of 0.164 and a standard deviation of 0.075. Clearly,
our observed eccentricity of $0.115\pm0.087$ does not rule out
a circular orbit.

Our $N=74$ observations of HD 179079 are spread nearly uniformly
in orbital phase and have an rms residual of $\sigma=3.9$ \ms\
(dominated by stellar jitter with an amplitude of about 3.5 \ms).
With a measured eccentricity of $e=0.11$ and a velocity
semiamplitude of $K=6.6$ \ms, we obtain $e/\sigma_e=1.1$,
which is well below the approximate threshold for a significant
detection.

The dotted line in Figure \ref{fig_ebias} shows the predicted
distribution according to Equation \ref{eq_edist}, but with
$\sigma=4.2$ \ms, rather than the observed rms of 3.9 \ms,
to better match the observed distribution. This slight excess
in $\sigma_e$ relative to Equation \ref{eq_sigmae} may be due
to velocity constraints that are not uniformly distributed
in phase or non-optimal behavior of the Levenberg-Marquardt
algorithm used to fit Keplerian orbits. The Keplerian fitting
routine also returns the limiting value of $e=0$ more often
than predicted by the analytic approximation.

\subsection{Observed Eccentricity Distributions}

\begin{figure}
\plotone{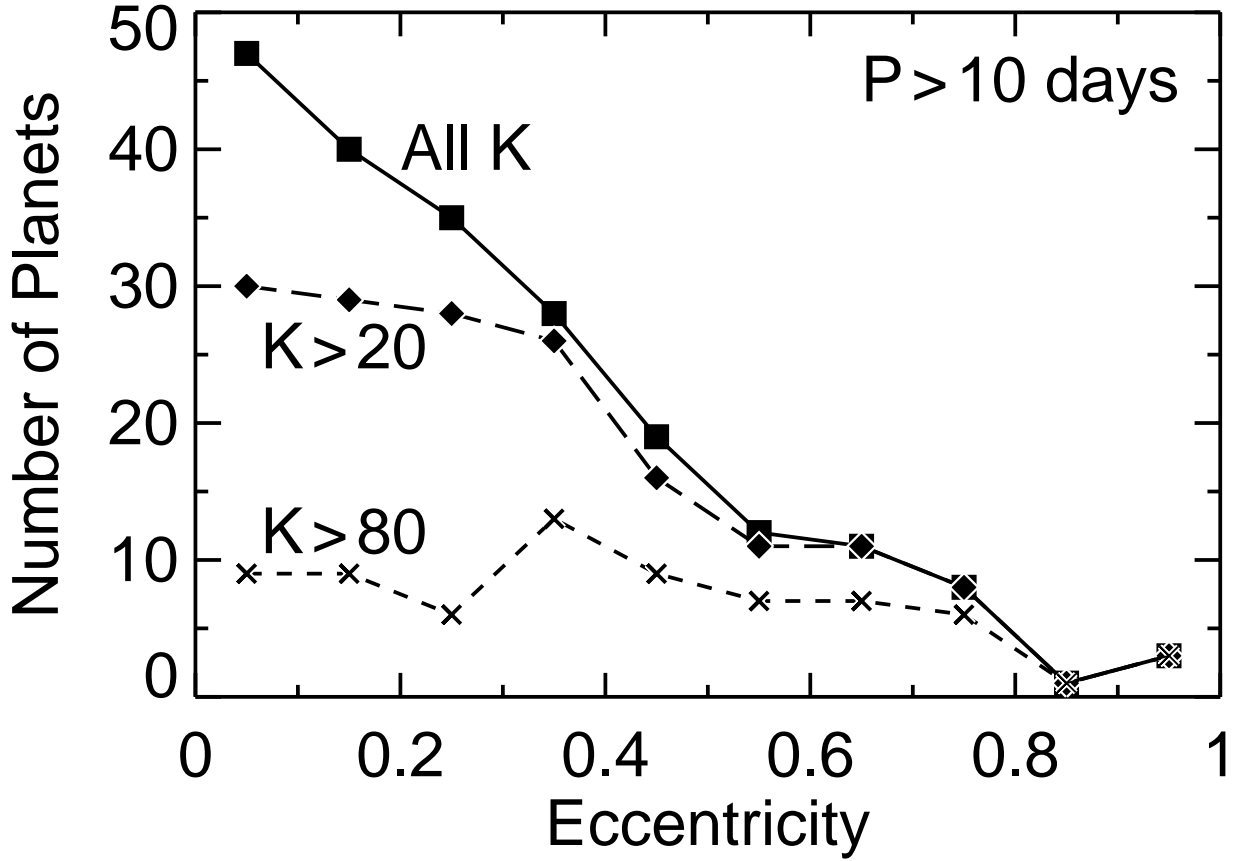}
\figcaption{
Ignoring planets with periods shorter than 10 days, which may
have circularized, the eccentricity distribution for known
planets decreases linearly from $e=0$ to $e=0.6$. For planets
with $K>20$ \ms, the distribution is much flatter from $e=0$
to $e=0.4$ and then declines. For planets with $K>80$ \ms,
the distribution is flat all the way up to $e=0.8$. These
distributions may reflect an observational bias against
detecting high eccentricity planets with velocity
semiamplitudes $K$ near the detection limit, in which
case the observed distribution for large $K$ may represent
the intrinsic eccentricity distribution for giant planets.
\label{fig_ecc_dist}}
\end{figure}

Figure \ref{fig_ecc_dist} shows observed eccentricity
distributions for 204 well-characterized planets, a subset
of 163 planets with $K>20$ \ms, and a further subset of 70
planets with $K>80$ \ms. In all three cases, we exclude 53
planets with periods less than 10 days that may have
experienced orbital evolution due to tidal interaction
with the star.

The observed eccentricity distributions become flatter as
lower $K$ planets are excluded. This flattening cannot be
due to eccentricity measurement bias (Section 5.1),
which would create the opposite slope, as positively
biased measurements for lower $K$ planets are included
in the distribution.
The sequence of observed eccentricity distributions in
Figure \ref{fig_ecc_dist} may reflect dynamical processes
that link planet mass and eccentricity \citep[e.g.,][for
high mass planets]{jur08,for08}, but below we discuss
briefly a possible origin based on observational bias.

At fixed $K$, planets in more eccentric orbits induce large
velocity shifts for a smaller fraction of the orbit, making
them harder to detect. For this reason, the dependence of
the observed eccentricity distribution on $K$ threshold may
reflect an observational bias, rather than measurement bias.
If true, the flat eccentricity distribution for $K>80$ \ms\
in Figure \ref{fig_ecc_dist} may be the true distribution for
giant planets. For the $K>20$ \ms\ threshold, detection
of highly eccentric planets may become more difficult,
causing the eccentricity distribution to drop above $e=0.4$.
Finally, the lowest $K$ planets may be very difficult to
detect reliably except in nearly circular orbits.
If this \textit{hypothesis} is true, a large population
of low $K$, high $e$ planets have yet to be detected.
It would be interesting to quantify the expected
detection bias as a function of eccentricity in actual
radial velocity planet search programs.

\section{Discussion}

We have presented two exoplanets detected at Keck Observatory in
a search for hot Neptunes and other low-amplitude planets. The
two planets have masses $1.61\pm0.15$ and $20.7\pm2.2$ times the
mass of Neptune. The radial velocity semiamplitudes of these
planets are $6.64\pm0.60$ and $16.3\pm1.3$ \ms, respectively.
Both host stars are metal rich subgiants, which is consistent
with the selection criteria of the parent N2K sample
\citep{fis05a}. 

HD 179079 is a G5 subgiant with an  $\msini=27.5\pm2.5$ \mear\
planet in a low eccentricity, 14.48 d orbit. The semivelocity
amplitude of the star is only 6.6 \ms, making this a challenging
detection. Although the periodogram signal was apparent after
about 30 observations, we obtained 74 Doppler measurements before
announcing this planet because of the small velocity amplitude
and concern about jitter in this slightly evolved star. However,
the periodogram power continues to grow at the same period and
the rotational period for this subgiant is expected to be about
38 days. 

HD 73534b is a $\msini=1.103\pm0.087$ \mjup\ planet in a nearly
circular 4.85 yr orbit around a metal-rich subgiant. This is
one of the handful of known planets \citep[e.g.,][]{wri08} in low
low eccentricity orbits that did not migrate well inside the
ice line in the parent protoplanetary disk. As precise Doppler 
observations extend over longer time baselines, more such planets 
will be discovered.

Numerical simulations of the efficiency of orbital migration
\citep[][Figure 5]{ida04,ida08} provide predictions for the mass
and semi-major axis distribution of exoplanets. The distribution
of gas giant planets provides a good match to the numerical
prescriptions, however the most remarkable feature of the
simulations is the prediction of a planet ``desert.'' 
Over a wide range of initial conditions (migration speeds,
stellar metallicity, and mass or surface density in the
protoplanetary disks), these results consistently show a paucity
of intermediate mass ($M_p \sim 10 - 100$ \mear) planets closer
than a few AU. One of the two planets presented here, HD 179079,
populates this ``no planet'' region, providing an interesting,
benchmark for future simulations. 

\begin{figure}
\plotone{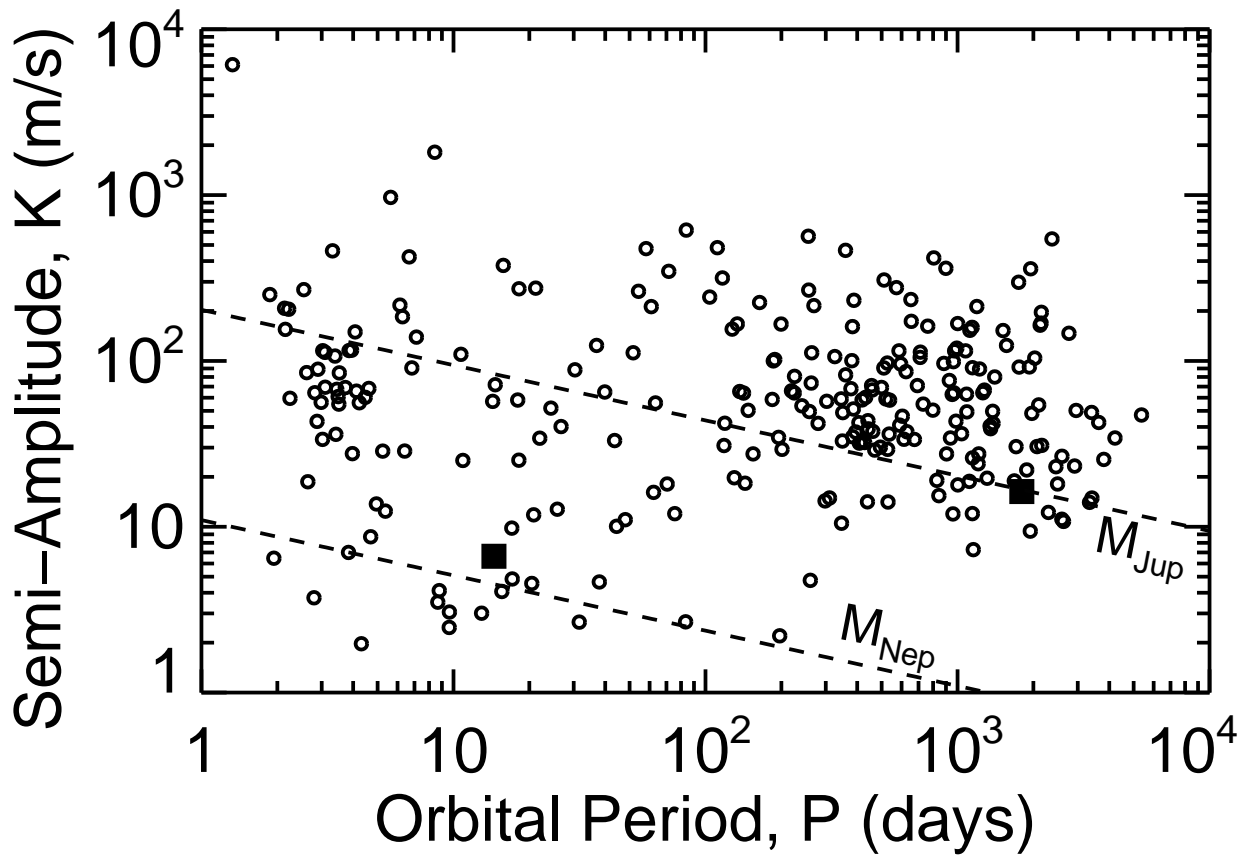}
\figcaption{Velocity semiamplitude vs. orbital period
for planets with radial velocity detections. The two
planets announced in this paper are indicated by filled
squares. Diagonal dashed lines show the approximate
locus of Jupiter and Neptune mass planets, assuming
$M=\msun$, $e=1$, and $\sin i=1$. Detecting low mass
planets with long periods is particularly challenging.
\label{fig_rv_detection}}
\end{figure}

The detected stellar velocity amplitudes are small for both new
planets. Figure \ref{fig_rv_detection} compares the masses and
orbital periods of these new planets with 256 known exoplanets.
The Doppler technique has progressed both in precision and
duration to a point where the detection of planets with amplitudes
of 5 \ms\ define a boundary for orbital periods out to 10 years.
Clearly, we stand at the threshold for detecting the signpost
of our solar system, Jupiter, with a 12 \ms\ velocity amplitude
in a 11.7 year orbit.

\acknowledgements
We gratefully acknowledge the dedication and support of the Keck
Observatory staff, in particular Grant Hill for support with HIRES.
D.A.F. acknowledges research support from NASA grant NNX08AF42G.
G.W.H. acknowledges support from NASA, NSF, Tennessee State University,
and the State of Tennessee through its Centers of Excellence program.
We thank the NASA Exoplanet Science Institute (NExScI) for 
support through the KPDA program. We thank the NASA and NOAO 
Telescope assignment committees for allocations of telescope time.  
The authors extend thanks to those of Hawaiian ancestry on whose 
sacred mountain of Mauna Kea we are privileged to be guests.  
Without their kind hospitality, the Keck observations presented 
here would not have been possible. This research has made use of 
the SIMBAD database, operated at CDS, Strasbourg, France, and of 
NASA's Astrophysics Data System Bibliographic Services.

\clearpage
\clearpage

\clearpage
\clearpage

\clearpage
\clearpage

\clearpage
    
\clearpage

\clearpage

\clearpage

\clearpage

\begin{deluxetable}{lll}
\tablecaption{Stellar Parameters\label{tab_stellar_params}}
\tablewidth{0pt}
\tablehead{
 \multicolumn{1}{l}{Parameter} &
 \multicolumn{1}{l}{HD 179079} &
 \multicolumn{1}{l}{HD 73534}
}
\startdata
Spectral type   & G5 IV        & G5 IV       \\
Distance (pc)   & 65.5(3.3)    & 81.0(4.9)   \\
$V$             & 7.95         & 8.23        \\
$B-V$           & 0.744(13)    & 0.962(21)   \\[6pt]
\teff\ (K)      & 5684(44)     & 5041(44)    \\
\logg           & 4.062(60)    & 3.780(60)   \\
\fe             & 0.250(30)    & 0.232(30)   \\
\vsini\ (\kms)  & $<1.0$       & $<1.0$      \\[6pt]
\bc             & $-$0.094     & $-$0.266    \\
\mbol           & 3.77         & 3.42        \\
\lstar\ (\lsun) & 2.41(27)     & 3.33(43)    \\
\rstar\ (\rsun) & 1.599(92)    & 2.39(16)    \\
\mstar\ (\msun) & 1.146(28)    & 1.228(60)   \\[6pt]
\shk            & 0.153        & 0.155       \\
\rhk            & $-$5.06      & $-$5.13     \\
\prot\ (d)      & $\sim38$     & $\sim53$
\enddata
\tablecomments{Parentheses after each table entry enclose the
uncertainty in the last two or three tabulated digits. For
example, 0.744(13) is equivalent to $0.744\pm0.013$ and
81.0(4.9) is equivalent to $81.0\pm4.9$.} 
\end{deluxetable}

\clearpage

\begin{deluxetable}{lrrr}
\tablecaption{HD 179079 Parameters versus Iteration\label{tab_179079_iter}}
\tablewidth{0pt}
\tablehead{
   \multicolumn{1}{l}{Parameter}
 & \multicolumn{1}{r}{Iter 1}
 & \multicolumn{1}{r}{Iter 2}
 & \multicolumn{1}{r}{Iter 3}
}
\startdata
\rchisq         &     8.60 &     9.17 &     9.18 \\
\logg\ [SME]    &    4.107 &    4.074 &    4.063 \\
\logg\ [Iso]    &    4.074 &    4.063 &    4.062 \\
\teff\ (K)      &     5692 &     5686 &     5683 \\
\fe             &    0.267 &    0.250 &    0.250 \\
\bc             & $-0.092$ & $-0.094$ & $-0.094$ \\
\lstar\ (\lsun) &    2.404 &    2.407 &    2.408 \\
\rstar\ (\rsun) &    1.593 &    1.597 &    1.599 \\
\mstar\ (\msun) &    1.153 &    1.147 &    1.146 \\
\enddata
\end{deluxetable}

\clearpage

\begin{deluxetable}{rrc}
\tablecaption{Radial Velocities for HD 179079\label{tab_rv_179079}}
\tablewidth{0pt}
\tablehead{
 \colhead{} &
 \colhead{RV} &
 \colhead{$\sigma_{\rm RV}$}   \\
  \colhead{JD-2440000} &
  \colhead{(\ms)} &
  \colhead{(\ms)}
}
\startdata
    13197.99712  &      5.70  &      2.21   \\ 
    13198.96331  &      8.79  &      2.96   \\ 
    13199.90927  &      2.30  &      2.45   \\ 
    13208.02469  &    -13.65  &      2.10   \\ 
    13603.86010  &      7.88  &      1.43   \\ 
    13961.87556  &     -9.86  &      1.45   \\ 
    13963.86704  &     -2.02  &      1.51   \\ 
    13981.76031  &      3.64  &      1.34   \\ 
    13982.80620  &      4.89  &      1.36   \\ 
    13983.77032  &      2.27  &      1.29   \\ 
    13984.84523  &     -2.68  &      1.27   \\ 
    14249.03715  &    -13.04  &      1.32   \\ 
    14250.08001  &    -12.97  &      1.40   \\ 
    14251.05687  &     -6.73  &      0.99   \\ 
    14251.93641  &     -1.91  &      1.07   \\ 
    14256.08978  &      6.74  &      1.24   \\ 
    14279.03934  &     -6.23  &      1.10   \\ 
    14280.04708  &    -11.07  &      1.17   \\ 
    14286.03795  &     -1.14  &      1.33   \\ 
    14304.97219  &     -4.18  &      1.26   \\ 
    14305.97242  &     -7.59  &      1.15   \\ 
    14306.97185  &     -1.31  &      1.12   \\ 
    14308.00091  &     -1.65  &      1.18   \\ 
    14308.96870  &      1.09  &      1.12   \\ 
    14309.96526  &     -2.20  &      1.18   \\ 
    14310.95716  &      4.63  &      1.04   \\ 
    14311.95490  &      4.60  &      1.22   \\ 
    14312.95049  &      3.91  &      1.27   \\ 
    14313.94773  &      5.53  &      1.14   \\ 
    14314.95737  &     13.65  &      1.06   \\ 
    14318.86262  &     -7.74  &      1.09   \\ 
    14335.96501  &     -3.14  &      1.10   \\ 
    14336.98916  &      3.24  &      1.23   \\ 
    14339.85272  &      6.09  &      1.03   \\ 
    14343.88818  &      1.41  &      1.21   \\ 
    14344.94423  &     -1.79  &      1.15   \\ 
    14345.75855  &     -5.08  &      1.16   \\ 
    14396.72957  &     -1.40  &      1.22   \\ 
    14397.75656  &     -2.19  &      1.14   \\ 
    14398.74164  &     -0.82  &      1.12   \\ 
    14399.72544  &     -1.40  &      1.40   \\ 
    14427.74492  &      1.30  &      1.21   \\ 
    14428.70443  &      3.35  &      1.22   \\ 
    14429.68634  &      3.74  &      1.24   \\ 
    14430.68308  &     -1.72  &      1.20   \\ 
    14548.15175  &     -1.22  &      1.19   \\ 
    14549.14521  &     -6.58  &      1.88   \\ 
    14602.97662  &      3.42  &      1.20   \\ 
    14603.99723  &     13.06  &      1.44   \\ 
    14634.04919  &     -0.11  &      1.36   \\ 
    14634.97773  &      0.10  &      1.33   \\ 
    14636.02087  &     -3.08  &      1.23   \\ 
    14637.06667  &     -7.61  &      1.33   \\ 
    14638.01419  &     -7.99  &      1.13   \\ 
    14639.04519  &     -8.91  &      1.20   \\ 
    14640.12738  &    -11.60  &      1.26   \\ 
    14641.00624  &     -9.48  &      1.11   \\ 
    14642.10448  &     -6.49  &      1.32   \\ 
    14644.10005  &     -1.80  &      1.24   \\ 
    14674.83794  &      6.63  &      1.19   \\ 
    14688.84946  &      4.80  &      1.32   \\ 
    14690.02171  &      5.29  &      1.34   \\ 
    14717.77181  &      2.51  &      1.20   \\ 
    14718.79085  &      4.81  &      1.14   \\ 
    14719.80332  &      4.79  &      1.09   \\ 
    14720.84228  &      5.45  &      1.18   \\ 
    14721.82896  &     -2.64  &      1.13   \\ 
    14722.77201  &      1.66  &      1.18   \\ 
    14723.76538  &     -3.10  &      1.27   \\ 
    14724.77867  &     -2.92  &      1.27   \\ 
    14725.76972  &    -10.69  &      1.25   \\ 
    14726.76598  &     -6.77  &      1.16   \\ 
    14727.84278  &    -11.92  &      1.16   \\ 
    14777.76256  &      7.17  &      1.25   \\ 
\enddata
\end{deluxetable}
\clearpage

\clearpage
 
\begin{deluxetable}{lll}
\tablecaption{Orbital Parameters\label{tab_orbital_params}}
\tablewidth{0pt}
\tablehead{
 \multicolumn{1}{l}{Parameter}  &
 \multicolumn{1}{l}{HD 179079b} &
 \multicolumn{1}{l}{HD 73534b}
} 
\startdata
$P$ (d)             & 14.476(11)     & 1770(40)     \\
$K$ (\ms)         & 6.64(60)       & 16.2(1.1)    \\
$e$               & 0.115(87)      & 0.074(71)    \\
\Tp\ (JD)         & 2454400.5(2.4) & 2456450(850) \\
$\omega$ (deg)    & 357(62)        & 12(66)       \\[6pt]
\msini\ (\mjup)   & 0.0866(80)     & 1.103(087)   \\
\msini\ (\mear)   & 27.5(2.5)      & 350(27)      \\
$a$ (AU)          & 0.1216(10)     & 3.067(68)    \\[6pt]
$N_{\rm obs}$     & 74             & 30           \\
Jitter (\ms)      & 3.5            & 3.5          \\
rms (\ms)         & 3.88           & 3.36         \\
\rchisq           & 1.04           & 0.91
\enddata                        
\tablecomments{Parentheses after each table entry enclose the
uncertainty in the last two or three tabulated digits. For
example, 14.476(11) is equivalent to $14.476\pm0.011$ and
2456450(850) is equivalent to $2456450\pm850$.} 
\end{deluxetable}                

\clearpage

\begin{deluxetable}{rrc}
\tablecaption{Radial Velocities for HD 73534\label{tab_rv_73534}}
\tablewidth{0pt}
\tablehead{
 \colhead{} &
 \colhead{RV} &
 \colhead{$\sigma_{\rm RV}$}
\\
 \colhead{JD - 2440000} &
 \colhead{(\ms)} &
 \colhead{(\ms)}
}
\startdata
    13014.92199  &     13.92  &      1.61   \\ 
    13015.91840  &     20.34  &      1.45   \\ 
    13016.92449  &     14.69  &      1.65   \\ 
    13071.88830  &     18.52  &      1.79   \\ 
    13369.04318  &     -4.59  &      1.04   \\ 
    13369.90891  &     -5.77  &      0.98   \\ 
    13397.90482  &      1.29  &      1.00   \\ 
    13746.93249  &     -9.00  &      1.13   \\ 
    13747.97716  &     -6.34  &      1.11   \\ 
    13748.93288  &    -10.62  &      1.14   \\ 
    13749.86191  &    -13.88  &      1.03   \\ 
    13750.86815  &    -20.34  &      1.15   \\ 
    13752.97936  &     -9.79  &      0.92   \\ 
    13775.94527  &     -9.63  &      1.01   \\ 
    13776.83750  &    -15.35  &      1.21   \\ 
    14130.09538  &      1.43  &      1.29   \\ 
    14428.05313  &     13.21  &      0.98   \\ 
    14428.99770  &     18.79  &      1.00   \\ 
    14779.07344  &     18.32  &      1.05   \\ 
    14780.12264  &     18.86  &      1.09   \\ 
\enddata
\end{deluxetable}
\clearpage

\clearpage 

\begin{deluxetable}{lrrrrrrr}
\tablecaption{HD 73534 Parameters versus Iteration\label{tab_73534_iter}}
\tablewidth{0pt}
\tablehead{
   \multicolumn{1}{l}{Parameter}
 & \multicolumn{1}{r}{Iter 1}
 & \multicolumn{1}{r}{Iter 2}
 & \multicolumn{1}{r}{Iter 3}
 & \multicolumn{1}{r}{Iter 4}
 & \multicolumn{1}{r}{Iter 5}
 & \multicolumn{1}{r}{Iter 6}
 & \multicolumn{1}{r}{Iter 7}
}
\startdata
\rchisq         &    43.91 &    44.40 &    45.11 &    45.57 &    45.78 &    45.86 &    46.10 \\
\logg\ [SME]    &    3.576 &    3.709 &    3.746 &    3.767 &    3.770 &    3.777 &    3.780 \\
\logg\ [Iso]    &    3.709 &    3.746 &    3.767 &    3.770 &    3.777 &    3.780 &    3.779 \\
\teff\ (K)      &     4946 &     4990 &     5018 &     5023 &     5034 &     5040 &     5038 \\
\fe             &    0.165 &    0.201 &    0.225 &    0.228 &    0.235 &    0.232 &    0.234 \\
\bc             & $-0.304$ & $-0.284$ & $-0.273$ & $-0.272$ & $-0.268$ & $-0.266$ & $-0.267$ \\
\lstar\ (\lsun) &    3.451 &    3.389 &    3.356 &    3.351 &    3.338 &    3.332 &    3.335 \\
\rstar\ (\rsun) &    2.528 &    2.461 &    2.422 &    2.415 &    2.400 &    2.392 &    2.396 \\
\mstar\ (\msun) &    1.178 &    1.209 &    1.224 &    1.226 &    1.229 &    1.228 &    1.229 \\
\enddata
\end{deluxetable}

\clearpage

\end{document}